\documentclass[
    ,final            
  ]
  {aipproc}

\layoutstyle{6x9}

\begin{document}

\title{Status of GRB Observations with the Suzaku  Wideband All-sky Monitor}


\author{M. S. Tashiro, Y. Terada, Y. Urata, K. Onda, N. Kodaka, A. Endo, M. Suzuki, K. Morigami}{
  address={Saitama University, 255 Shimo-Okubo, Sakura, Saitama, 338-8570, Japan}
}

\author{K. Yamaoka, Y. E. Nakagawa, S. Sugita}{
  address={Aoyama Gakuin University, 5-10-1 Fuchinobe, Sagamihara, 229-8558, Japan}
}

\author{Y. Fukazawa, M. Ohno, T. Takahashi, C. Kira, T. Uehara}{
  address={Hiroshima University, 1-3-1 Kagamiyama, Higashi-Hiroshima, 739-8526, Japan}
}

\author{T. Tamagawa}{
  address={RIKEN, 2-1 Hirosawa, Wako, 351-0198, Japan}
}

\author{T. Enoto, R. Miyawaki, K. Nakazawa, K. Makishima}{
  address={University of Tokyo, 7-3-1 Hongo, Bunkyo, Tokyo, 113-0033, Japan}
}

\author{E. Sonoda, M. Yamauchi, S. Maeno, H. Tanaka, R. Hara}{
  address={University of Miyazaki,  Gakuen-kibanadai, Miyazaki, 889-2192, Japan}
}

\author{M. Suzuki, M. Kokubun, T. Takahashi}{
  address={ISAS/JAXA, 3-1-1 Yoshinodai, Sagamihara, 229-8510, Japan}
}
\author{S. J. Hong}{
  address={Nihon University, 7-24-1Narashinodai, Funabashi, 274-8501, Japan}
}
\author{T. Murakami}{
  address={Kanazawa University, Kakuma-cho, Kanazawa, 920-1192, Japan}
}
\author{H. Tajima}{
  address={Stanford Linear Accelerator Center, 2575 Sand Hill Road M/S 29 Menlo Park, CA 94025, U.S.A. }
}

\begin{abstract}
The Wide-band All-sky Monitor (WAM) is a function of the large lateral BGO shield of the Hard X-ray Detector (HXD) onboard Suzaku. Its large geometrical area of 800 cm$^2$ per side, the large stopping power for the hard X-rays and the wide-field of view make the WAM an ideal detector for gamma-ray bursts (GRBs) observations in the energy range of 50-5000 keV. In fact, the WAM has observed 288 GRBs confirmed by other satellites, till the end of May 2007.
\end{abstract}

\maketitle


\section{Suzaku and the HXD/WAM}

The fifth Japanese cosmic X-ray satellite Suzaku\cite{mitsuda:07} launched on 10 July, 2005 is operated in the low earth orbit with an altitude of 570 km and the inclination angle of $31^\circ$. Suzaku covers a broad energy range 0.2 -- 600 keV with the two kinds of instruments: the four X-ray CCDs (XISs) installed on the focal plane of the X-ray Telescopes (XRT) and the Hard X-ray Detector (HXD).
The HXD consists of 16 well-type GSO-BGO phoswitch counters and surrounding 20 BGO active shield counters\cite{takahashi:07}.
The large and thick BGO crystal realizes not only effective passive shields but also active shield with the compound eye configurations of the HXD.
At the same time, these active shields are also utilized as the wide-band all-sky monitor (WAM) \cite{yamaoka:05}\cite{ohno:05}\cite{terada:05}\cite{hong:05} in soft gamma-ray band.

The WAM has a geometrical area of 800 cm$^2$ in each side, and the large stopping power of the BGO crystal realizes an area of 400 cm$^2$ even for the 1 MeV photons as shown in Fig.~1.
The larger effective area in 400 keV to 5 MeV band than the other gamma-ray burst (GRB) detectors enables us to study high energy emissions from GRBs up to the MeV range. The WAM is one of ideal instruments to investigate the GRB peak energy  ($E_{\rm peak}$) distributing above several hundreds keVs.

\begin{figure}
  \includegraphics[height=.22\textheight]{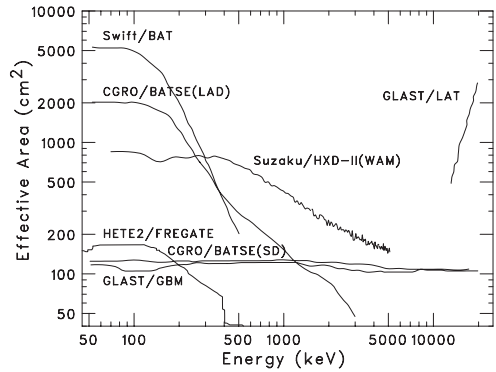}
  \caption{Effective areas of the GRB detectors in orbits. The effective areas of the WAM, the BATSE/LAD, the Swift/BAT, the HETE2/FREGATE, and the GLAST/LAT are calculated for the full absorption peak efficiency, although cross sections for the Compton scattering are included only for the GLAST/GBM, and the BATSE/SD.}
\end{figure}

GRBs are triggered with a similar logic installed in the BATSE onboard Compton Gamma-ray Observatory. Comparing the 1/4-s and 1 s counting rate with the previous 8-s average background, the hardware triggers the GRB or other bright transient events with the 5.7 $\sigma$ threshold. Onboard software also monitors the count rate and judge the deviation from the average, although that function is not operated as the trigger so far.
No prompt response capability is equipped, but the triggered and stored GRB data are down linked only at the ground station passages of Suzaku.

The instrument is monitored and maintained by the WAM team, who also reviews the obtained data from GRBs, solar flares and other transient objects.
All the functions have been working well. Although the photomultiplier tubes gain decreased by $\sim 20$ \% through the two years, the degradation has been compensated by the amplifier gain correction by command.
The WAM also contributes to the Interplanetary Network (IPN) to determine the GRB positions with other GRB satellites and probes in the interplanetary space.

\begin{figure}[htb]
  \includegraphics[height=.22\textheight]{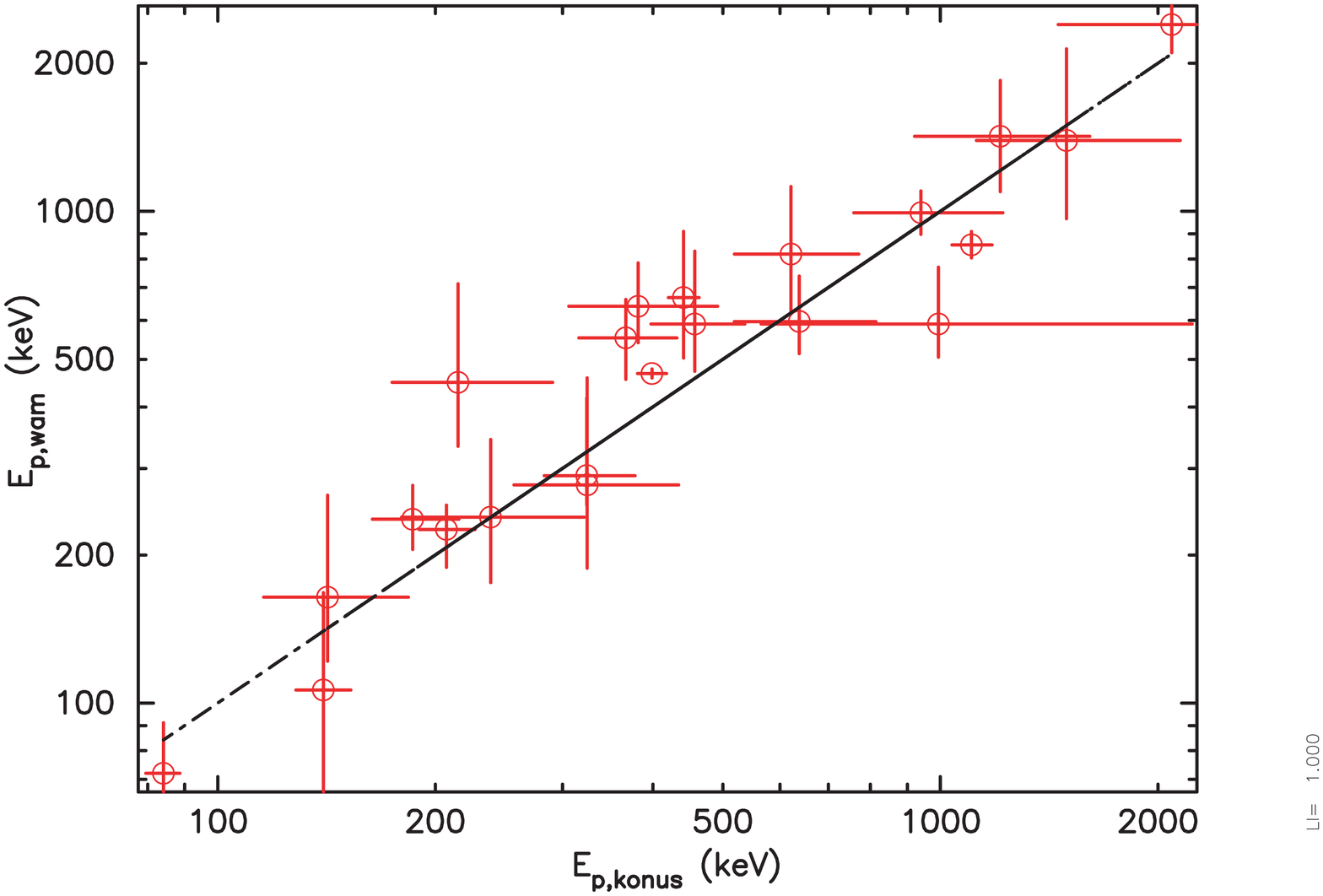}
  \caption{Comparison with the Konus-Wind results of the derived $E_{\rm peak}$ from simultaneously observed GRBs.}
\end{figure}

\section{Improvement of Energy Response}
The WAM is installed in the spacecraft and surrounded by other focal plane detectors, the structures and panels of the satellite.
Since the GRB irradiates the WAM through the surrounding materials, one need to calculate each effective area and energy response matrix reflecting their absorption in each direction.
The WAM energy response matrix is generated by a Geant4 based full simulation program.
We had revised the Suzaku mass model, reviewed the high-energy efficiency including Compton scattering above 5 MeV, and developed a new response generator. 
Utilizing some simultaneously observed GRB data, we carried out cross calibrations with the Konus-Wind. Figure~2 demonstrates the cross calibration results in the obtained $E_{\rm peak}$ with both instruments.

\section{OBSERVATION STATUS}
The WAM triggered over 700 events through 22 August 2005 to 31 August 2007, including 288 GRBs confirmed with one or more other satellites and 205 GRB-like events (table~1). We have reported more than 50 GCN circulars, including 29 WAM spectral analysis results during the same period.
The measured time durations of GRBs are summarized in Fig.~3, which still shows discrete distributions of the classical ``short'' and the ``long'' GRBs.

\begin{figure}[htb]
  \includegraphics[height=.22\textheight]{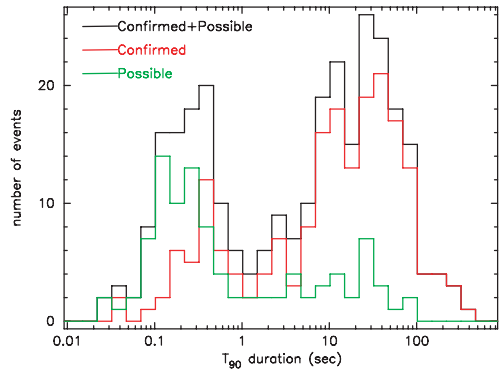}
  \caption{Time durations ($T_{90}$) distribution of the triggered GRBs by the WAM. Both ``confirmed'' and ``possible'' GRB events are indicated with the total (see text).}
\end{figure}

\begin{table}
\begin{tabular}{ccccc}
\hline
   \tablehead{1}{c}{b}{Confirmed GRB\tablenote{by simultaneous detection with other satellites.}}
  & \tablehead{1}{c}{b}{(localized)}
  & \tablehead{1}{r}{b}{Possible GRB}
  & \tablehead{1}{r}{b}{SGR\tablenote{soft gamma-ray repeater}}
  & \tablehead{1}{r}{b}{Solar Flare}   \\
\hline
 288 & (100) & 205 & 39 & 157\\
\hline
\end{tabular}
\caption{Statistics of the WAM triggered events, 
through 22 Aug. 2005 to 31 Aug. 2007.}
\end{table}


The processed data are available at the ISAS/JAXA web page\footnote{http://www.astro.isas.jaxa.jp/suzaku/HXD-WAM/}. Also recent results were seen in contribution papers in this volume by Ohno et al, Onda et al., Sugita et al., Uehara et al., Hurley et al., and Krimm et al.

\begin{theacknowledgments}
The authors thank the Suzaku team, the Konus-Wind team and the Swift/BAT team for their kind support in operation and the cross calibration. 
\end{theacknowledgments}



\bibliographystyle{aipproc}   

\bibliography{ms}

\IfFileExists{ms.bbl}{}
 {\typeout{}
  \typeout{******************************************}
  \typeout{** Please run "bibtex \jobname" to optain}
  \typeout{** the bibliography and then re-run LaTeX}
  \typeout{** twice to fix the references!}
  \typeout{******************************************}
  \typeout{}
 }

\end{document}